\documentstyle[prl,aps,floats]{revtex}
\begin{document}
 
\twocolumn[
\hsize\textwidth\columnwidth\hsize\csname@twocolumnfalse\endcsname
 
\title{Novel aspects of spin-polarized transport and spin dynamics}
 
\author{Igor \v{Z}uti\'{c}}
\address{Department of Physics, University of Maryland at  College
Park, College Park, Maryland 20742-4111, USA}
\maketitle

\begin{abstract}
There is a renewed interest to study spin-polarized transport and spin
dynamics in various electronic materials. The motivation to examine
the spin degrees of freedom (mostly in electrons, but also in 
holes and nuclei) comes from various sources: ranging from novel 
applications which are either not feasible or ineffective with 
conventional electronics, to using spin-dependent phenomena to 
explore the fundamental properties of solid state systems. Taken in a
broader context, term {\em spintronics} is addressing various aspects 
of these efforts and stimulating new interactions between different
subfields of condensed matter physics. Recent advances in material 
fabrication made it possible to introduce the nonequilibrium spin
in novel class of systems, including ferromagnetic 
semiconductors, high temperature superconductors and carbon 
nanotubes--which leads to a question of how such a spin could be utilized.
For this purpose it is important to extend the understanding of 
spin-polarized transport and spin dynamics to 
consider inhomogeneous systems, various heterostructures,
and the role of interfaces. This article presents some views on novel 
aspects of spin-polarized transport and spin dynamics (referring also 
to the topics which were addressed at the conference Spintronics 2001)
and suggests possible future research directions.
\end{abstract} 

]


The foundations for many aspects of spin-dependent phenomena 
which are currently investigated in the
context of electronic materials have 
been developed over the past several decades. One of the basic
ingredients--electrical spin injection (as a method to create
nonequilibrium spin population) in metals~\cite{aronov1},
semiconductors~\cite{aronov2}, and superconductors~\cite{aronov3},
has already been proposed by Aronov in the 70's. Other important examples, 
electrical 
detection of spin-polarized current through tunneling measurements
in ferromagnet/superconductor junctions\cite{tedrow71}, optical 
generation of spin-polarized carriers  
in semiconductors (spin pumping and optical 
orientation)~\cite{lampel68,parsons69,dyakonov71,optical84} 
studies~\cite{yafet52,elliott54,pines57,abrahams57,feher59,feher61,yafet63,gordon58,long} 
spin relaxation and spin coherence times (often denoted
as T$_1$~\cite{long}, and T$_2$, respectively), or Rashba spin-orbit 
coupling~\cite{rashba60,rashba84}, have also been known for quite some time. 
It is also interesting to note some early ideas, perhaps not appreciated
at the time. For example, Ref.~\cite{gordon58} mentions spin-based
memory, and what can be called Feher's effect~\cite{dassarma00a}
suggests how to purely electronically control nuclear spins~\cite{feher59b}, 
a desirable prospect in the current context of quantum information processing.
Nevertheless, experimental demonstrations of some of the desired 
effects, accompanied with substantial challenges, have only recently been 
realized.  This is well illustrated by the example of 
electrical spin injection from a metallic ferromagnet into a 
semiconductor. Such a spin injection, together with the Rashba spin-orbit 
coupling\cite{rashba60,rashba84} was an essential element for one of the early 
proposals for spintronic devices--spin field-effect 
transistor (spin-FET)~\cite{datta90,dassarma01,dassarma01b} by Datta and Das. 
While the electrical spin injection in metals~\cite{johnson85,johnson88},   
and even in high temperature superconductors, has been 
achieved~\cite{vasko97,dong97,wei99,yeh99}, 
spin injection from a metallic ferromagnet into a 
semiconductor proved to be more 
difficult~\cite{roukes,hammar99,comment1,comment2,reply,tang00,ploog01,bland01,isakovic01}.
The corresponding problem of spin injection across an interface between
a magnetic and a nonmagnetic material, or consequently the conversion of 
spin-polarized current into unpolarized current, has similarities with the
conversion of charge current between normal (N) and superconducting (S) 
regions~\cite{blonder82}. This analogy with the standard situation
for charge transport in N/S systems was first employed to spin injection
into a normal metal~\cite{vanson} and later extended to 
semiconductors~\cite{schmidt00}. It was suggested that for an effective
spin injection (in a diffusive transport regime) it is important to
reduce the conductivity mismatch between materials. Assumptions leading
to the concept of  conductivity mismatch are  better satisfied
for two metals~\cite{friso} than when at least one of the materials is a 
semiconductor.
For ferromagnetic metals, in contact with semiconductors, several 
complications can arise, including band bending,  
interface effects (spin-polarized transport influenced by
a formation of Schottky barrier\cite{bland01,isakovic01}), and the possibility 
of bipolar (involving electrons and holes) transport. Even in the absence 
of conductivity mismatch, for a semiconductor {\it p-n} junction, consisting of 
magnetic and nonmagnetic regions, there still could be no spin injection
at low applied bias\cite{zutic02}. The concept of 
conductivity mismatch has stimulated important theoretical work, including
a proposal that it can be eliminated by insertion of a tunneling 
contact~\cite{rashba00}. Several experimental groups have implemented this
idea and substantial spin injection has been reported 
even at room temperature~\cite{jonker01,hammar01} 
(a ferromagnetic tip can also serve as an efficient
spin injector~\cite{alvarado92,labella01}).
Related to the efforts concerning the spin-polarized transport in 
semiconductors, these advances could mark an important shift in 
interest--from asking how to inject spin; to questions about what 
could be done with the (nonequilibrium) spin in a semiconductor.

In comparison, considering a similar time period, since the proposal 
of spin-FET, there has been a very significant progress in the context 
of metallic systems (which include metallic ferromagnets, normal metals 
and insulators)\cite{parkin91,prinz95,gregg97,prinz98,magn2}. The understanding of 
spin-polarized transport 
and spin dynamics has reached a mature level and many questions about what 
can be done with injected spin have been answered. There are already 
successful spintronic applications utilizing magnetoresistance~\cite{prinz95,gregg97,prinz98,magn2,magn3,magn4,magn5,magn6,magn7,wolf00,wolf00b} 
(mostly giant and tunneling magnetoresistance (GMR,TMR)) including magnetic
read head, magnetic sensors and nonvolatile magnetic random access
memory (MRAM).
In contrast, very little is known about the spin-polarized transport
and spin dynamics in inhomogeneous semiconductors and their 
heterostructures. Ideas about possible device applications also
require substantial additional efforts since, unlike conventional 
semiconductor electronics, even the simplest structures are only
beginning to be understood. However, semiconductor 
spintronics\cite{dassarma00a,dassarma01,sham99,ostr99} offers several
important advantages such as versatility in doping and 
fabrication of various structures, signal amplification, electronically tunable
spin-orbit coupling, optical manipulation, bipolar transport, and simple 
integration with the dominant semiconductor technology, among others.
Materials advances, leading to the III-V ferromagnetic semiconductors,
such as (In,Mn)As~\cite{munekata89,ohno92}, 
and (Ga,Mn)As~\cite{ohno96,vanesch97} have resulted in the constantly growing
list of additional compounds 
(GeMn~\cite{park01}, CdMnGeP~\cite{CdMnGeP}, TiCoO~\cite{tico}
(Ga,Mn)N~\cite{GaMnN,room},
(Ga,Mn)P~\cite{GaMnP}) including support for possible
room temperature ferromagnetism
in (Ga,Mn)N~\cite{room} and (Ga,Mn)P\cite{privhebard}.
Typically, in these materials the spin-polarized carriers are holes with
faster spin relaxation rates (influenced by the strong  spin-orbit coupling), 
but is also possible to have spin-polarized electrons and 
ferromagnetism at lower, non-degenerate, levels of doping~\cite{privcat},
which may be desirable for potential applications. 
Several methods to influence ferromagnetism have been demonstrated, for
example, it can be optically induced~\cite{munekata97,munekata01}, or 
controlled with gate voltage~\cite{ohno00}
(which changes carrier concentration and consequently the Curie temperature
of a semiconductor).
More recent studies~\cite{privmunekata,privjonker} have shown 
that only a small fraction of gate voltage reported in~\cite{ohno00} 
is sufficient to control ferromagnetism.
A possible historical obstacle for faster progress concerning these materials 
is that the ferromagnetism and semiconductors were traditionally
studied separately, and that some of the similarities with an earlier class of 
ferromagnets~\cite{vonmolnar67,vonmolnar67b} or colossal magnetoresistive 
materials (CMR) were not sufficiently explored.
Theoretical studies of ferromagnetic semiconductors
~\cite{macdonald00,macdonald01,dietl00,hill01,hill02,mona01,bhat01,amit01,sham01,janko01,dietl01,erwin01} are facing many challenges, such as, 
understanding the origin of  ferromagnetism, their complex phase diagram
(typically, the insulating phase, corresponding to higher doping, 
is less understood), or the role of disorder. Some of the approaches are 
using concepts developed for  strongly correlated systems~\cite{amit01} 
and similarities with earlier magnetic semiconductors~\cite{furdyna88}.
There are additional difficulties
related to the lack of systematic transport measurements and large
uncertainties in the basic material parameters. Furthermore, 
these materials may not be a simple homogeneous systems and there are
complications in the efforts to interpret measurements. For example,
temperature dependent magnetization (which often departs significantly
from a Brillouin function) can sensitively depend on the
type of measurement performed on the same sample~\cite{mpkennett}.

Both ferromagnetic~\cite{ohno99} and semimagnetic 
semiconductors~\cite{oest99b,fiederling99,jonker00}
(in the presence of applied magnetic field) can be employed 
to electrically inject spins into a nonmagnetic semiconductor.
Together with optical orientation and spin 
pumping\cite{parsons69,dyakonov71,optical84,awschalom99},
this provides various methods to  explore how to use
the nonequilibrium spin in semiconductors.
In a theoretical proposal~\cite{dassarma00} to study spin-polarized
{\it p-n} junctions (which could serve as a building block
for all-semiconductor spin transistor), it was shown that, with an 
inhomogeneous doping, the  magnetization could increase in the interior of a
nonmagnetic material, away from the point of spin injection~\cite{zutic01}.
Carrier spin polarization can be  tuned with the applied bias (which
can extend the spin diffusion range).
By shining circularly polarized light {\it p-n} junction can serve
as spin-polarized solar battery~\cite{zutic01b} giving rise to spin
and charge currents, accompanied by net voltage. There are also
several other methods to generate spin currents in 
semiconductors~\cite{bhat00,gainchev01,fabian01}, and to realize 
a spin-polarized battery~\cite{gainchev01}. It was shown experimentally 
that there could be a large 
increase~\cite{malajovich01} in the spin injection efficiency if a 
spin-polarized {\it p-n} junction is used instead of a heterostructure.

For an inhomogeneously doped magnetic semiconductor, the
magnitude of Zeeman splitting  could change with position 
resulting  in an effective magnetic force which separates carriers of 
different spins (analogous to  a homogeneously doped sample in the presence of 
inhomogeneous magnetic field~\cite{fabian01,frustaglia01}).
A theoretical study~\cite{zutic02} of bipolar spin-polarized transport in 
inhomogeneous
magnetic semiconductors predicts an exponentially large  magnetoresistance,
resulting from the magnetic field dependence of the population of the
minority carriers (following Boltzmann statistics, for nondegenerate doping).
Another study, for a unipolar transport, governed by the majority carriers,
considered the effect of magnetoresistance in a transistor 
structure~\cite{flatte01}. 
To fully exploit the effects of exponential magnetoresistance,
which can be realized by considering magnetic/nonmagnetic
{\it p-n} junction~\cite{zutic02}, it would be desirable to seek materials with 
large g-factors (g$>$500 is reported at low 
temperatures~\cite{dietl94}) even at room temperature. 
Several interesting effects can occur if spin is electrically injected at
the $n$-terminal of such {\it p-n} junctions. Nonequilibrium spin can lead to a
charge current even at no applied bias, as well as to an open circuit voltage
which changes sign if the spin polarization of injected carriers is reversed
(similar effects of spin-charge coupling were previously discussed in the
context of metallic systems~\cite{johnson85,johnson87}). 
There is also an associated
GMR effect, governed by the relative orientations of the 
spin polarization of the injected spin and that in the magnetic $p$-region.
For a unipolar transport, similar to  metallic systems, GMR effect
was  experimentally demonstrated~\cite{schmidt01}
in the all-semiconductor heterostructure where spin was electrically injected 
from a magnetic into a  nonmagnetic semiconductor.  
Magnetic/nomagnetic {\it p-n} junction could be also used to probe 
electronically
the spin-relaxation time (or consequently the spin diffusion length) 
by measuring current-voltage characteristics as a 
function of magnetic field~\cite{zutic02}. 
Several experimental studies have used similar inhomogeneous doping in magnetic
structures~\cite{roukes01,samarth01}, considering, for example, reverse bias and 
tunneling 
in ferromagnetic/nonmagnetic diodes~\cite{halperin01,kohda01} or diodes
consisting of both hole- and electron-doped manganites~\cite{mitra01}.
Bipolar spin-polarized transport~\cite{zutic02,zutic01} can also be 
extended to consider various implementations of all-semiconductor  
spin transistors~\cite{fabian02b}. The nonequilibrium spin could be 
electrically injected or generated by shining circularly polarized light,
and such devices would be capable, analogous to conventional
bipolar transistors, of amplifying signals. 

The efforts to experimentally demonstrate the  proposal of 
Datta and Das~\cite{datta90} were accompanied by 
substantial
difficulties, 
but studies of spin-polarized transport in hybrid ferromagnet/semiconductor 
structures, related to spin-FET geometry,  are still 
continuing~\cite{gardelis99,seba01,hu01,hu01b,grundler01,nitta01,nikolic01}.
In the ballistic limit it is convenient to use the simple approach
applied to the N/S systems~\cite{blonder82} and to extend it to spin-polarized
transport in semiconductors~\cite{zutic99b},
with interfacial scattering modeled by a 
$\delta$-function~\cite{hu01b,grundler01,nitta01}; however it is suggested
that to include Rashba spin-orbit coupling some care is needed~\cite{kamp01}.
Hybrid ferromagnet structures were also considered as a method
to combine the effects of magnetoresistance, extensively studied in metallic
systems, with the versatility of semiconductors~\cite{johnson98}. 
To overcome the problems of spin-injection, across the 
Schottky barrier, at a ferromagnet/semiconductor interface, in  several 
geometries hot-electron transport is used~\cite{monsma95,filipe98}
for different realizations of transistor devices 
(hot-electron transport in metallic systems can be used to perform a
ballistic electron magnetic microscopy~\cite{beem}). 
More recently, a magnetic tunnel transistor~\cite{parkin01a,parkin01b},
which can operate at room temperature (and with higher output currents,
than in the previous proposals~\cite{monsma95,filipe98}),
was examined. It combines the usual structure of a magnetic tunnel 
junction (MTJ), which serves as an emitter--base part of a transistor, 
together with a GaAs collector (which forms a Schottky barrier at the
interface with the ferromagnetic base). The degree of spin-filtering 
in MTJ, and consequently the magnitude of the collector current,
depends on the relative orientation between the two  
directions of magnetization in the emitter and base.
The magneto-collector current (defined analogously to GMR),
is nonmonotonic as a function of energy (controlled by the
emitter-base voltage) of injected hot-electrons, and can
attain up to 64\% (with output currents of 1 $\mu$A 
at room temperature and applied bias of 1.4 V). A possible
application of such magnetic tunneling transistor could be as an 
effective injector of spin-polarized current\cite{parkin01a}
into a semiconductor. By  applying a small
magnetic field ($\sim$15 Oe, is sufficient to switch the direction of 
magnetization in the base) the spin polarization of injected
current can be significantly changed.

There is a wide range of interest to study spin-polarized transport 
and spin dynamics
in structures which involve superconducting material. For example,
spin-dependent response of a superconductor can be used to probe the
unconventional pairing symmetries (to identify the orbital,
and  
the spin symmetry of the superconducting state). Alternatively, the superconducting
region can serve as a diagnostic tool for studying spin-polarized transport
in other materials or to investigate magnetically active interfaces.
While the  pioneering experiments on spin-dependent tunneling~\cite{tedrow71}
involving superconductors have influenced the early development
of spin-polarized
transport, perhaps two other aspects, involving superconductors, have 
generated a substantial interest in recent research. One of them pertains
to the materials advances and fabricating CMR/high temperature 
superconductor (HTSC) heterostructures. Related 
experiments~\cite{vasko97,dong97,wei99,yeh99,stroud98} and the spin injection 
into HTSC's made it possible to consider low-power 
superconducting devices which could have current gain and serve as very fast 
switches
(an alternative proposal, with heterostructures consisting of ferromagnetic 
semiconductors and superconductors, has been  suggested for implementation of 
logic circuits and switches~\cite{kulic00}).
However, theoretical
understanding of the corresponding nonequilibrium processes including
spin injection and spin relaxation, not limited to simple ferromagnets and
$s$-wave superconductors~\cite{zhao95,maekawa99} is still a formidable 
challenge facing
complex properties of both CMR's and HTCS's.
For example, even for a simple material such as Aluminum, 
a long-term puzzle related to spin relaxation has only recently been 
resolved~\cite{jaro98,jaro99}.
The other aspect, which has often been examined, is considering 
how the process of Andreev reflection
(in addition to  Andreev~\cite{andreev64} addressed independently
in a less known work of de Gennes and Saint 
James~\cite{degennes63})  
is influenced by 
spin polarization in both 
conventional~\cite{zutic99b,dejong95,soulen98,uphaday98,jedema99,melin00,mazin01,valls01}, 
and unconventional 
superconductors~\cite{zhu99,zutic99,kashiwaya99,si99,zutic00}. Andreev reflection 
in low temperature, $s$-wave, superconductors has already been established as a 
sensitive method to experimentally measure carrier spin polarization in various 
materials~\cite{soulen98,uphaday98,nadgorny01,ji01}.
If, on the other hand,
HTCS's~\cite{vasko98,chen01,dong01,kraus01} are considered, there is a much bigger
temperature range available to study spin-polarized transport across  interfaces with
magnetic materials. The sign change of the pair potential in unconventional
superconductors can give rise to Andreev bound state 
(midgap states)~\cite{hu94,kashiwaya95,wei98} and, as a consequence, to zero bias
conductance peak. In the presence of spin-polarization such Andreev bound
states can be suppressed~\cite{zutic99} and used in 
conductance measurements to estimate
the temperature dependence of an interfacial spin polarization at the
CMR/HTSC's interface~\cite{chen01} and examine the influence of interface
morphology on spin-polarized transport.  
In addition to ferromagnet/superconductor (F/S) junction, there is growing
interest to examine other configurations where spin-polarized Andreev
reflection plays an important role. For example, it was proposed that
the ferromagnetic contacts placed on the same side of superconductor
can lead to large magnetoresistive effects~\cite{deutscher00} and an
experimental realization could employ a geometry where one these ferromagnets
would be a spin-polarized STM tip~\cite{donchev01}. Some similar aspects
of modified Andreev reflection, which could arise when the two ferromagnetic
regions are on the same sides of a superconductor, were also examined 
with the addition of a quantum dot~\cite{zhu01} or by considering  
effects of metallic and insulating ferromagnets~\cite{melin01}.
Since the Josephson effect,
in the nonmagnetic systems, is related to Andreev reflection~\cite{furusaki91},
it is important to understand the effects of spin polarization on Andreev 
reflection for Josephson effect in S/F/S structures~\cite{bourgeois01}.
Andreev reflection can also serve to investigate entangled spin states
of electrons~\cite{recher01} 
(the wave function of two electrons is not a simple product of the  
wave functions for
individual electrons), an important concept for spin-based
quantum computing and quantum communication~\cite{dassarma00a,loss98,kane98,hu01,hu01b}.
In an earlier theoretical proposal to study spin-entanglement 
through transport measurements, geometries involving beam splitters and double 
quantum dot have been considered~\cite{loss1,loss2}. Since Cooper pairs 
are a natural source of  spin-entangled electrons it was 
suggested~\cite{dassarma00a} that they could be used in a similar transport 
measurements, where it may be even possible to consider different
signature of spin-triplet
pairing, the symmetry supported in the context of Sr$_2$RuO$_4$~\cite{maeno98}
and some quasi-one-dimensional superconductors~\cite{ishiguro98,organic01}.
However, additional complications would arise in systems with
Cooper pairs of mixed spin symmetry~\cite{gorkov01}.

One of the consequences, resulting from the studies of spin-polarized
transport and spin dynamics, is that it is often difficult
to limit contributions to a particular field. Instead, concepts
originally introduced in a specific context keep reappearing in 
seemingly unrelated areas and merging both fundamental and
applied aspects. In addition to studies of superconductivity,
systems of reduced dimensionality (quantum wires, quantum dots, etc.)
offer a similar example. With the possibility of fabricating carbon 
nanotubes~\cite{dekker99} as a natural realization of one dimensional (1D)
quantum wires, it is important to consider the influence of interactions
that could lead to the breakdown of Fermi-liquid theory
and to spin-charge separation (for example, spin and charge
excitations can have different velocities) in a 1D Luttinger liquid~\cite{voit95}. 
Spin-charge separation was also proposed in the context of HTSC's~\cite{anderson}
and there are suggestions of how spin injection~\cite{johnson85} and
transport measurement could be used to detect it in a Luttinger 
liquid~\cite{si98} and in HTSC's~\cite{si97}. Theoretical studies
of spin transport in Luttinger liquids~\cite{balents00,balents01} 
indicate that it is qualitatively different from the spin transport
in Fermi liquid (as well as from the charge transport in Luttinger
liquid). On one hand, this poses an additional difficulty to 
use experience from the research on spin-polarized transport
in more conventional electronic materials, while, on the other hand, 
supported by the findings for charge transport in carbon 
nanotubes~\cite{fara,marc,leonard00,odintsov00,fuhrer00},
it opens a possibility for molecular spintronics.
There are still only a small number of experiments which examine
the spin-dependent transport in carbon nanotubes~\cite{tsukagoshi,marcpriv},
while  some theoretical studies consider spin and charge 
pumping~\cite{chamon01}
and the effect of Rashba interaction~\cite{rashba60,rashba84}
in a Luttinger liquid~\cite{hausler}. It is also interesting to 
note~\cite{balents01}
the similarity of spin-dependent transport in a Luttinger liquid to
Andreev reflection in superconductor/normal metal/superconductor junctions. 
It is conceivable that, in addition to the
carbon nanotube analogue of magnetic tunnel junction~\cite{tsukagoshi},
there will soon be other configurations (heterojunctions, transistors,
etc.) that would explore spin-dependent transport and spin dynamics.
Perhaps, one of the future challenges could be to involve more 
carbon in spintronics and consider how  to
combine the spin degrees of freedom with the progress in C$_{60}$
superconductivity~\cite{batlogg}. 
 
\vspace{0.2cm} 
I would like to thank J. Fabian, S. Das Sarma, E. I. Rashba, and R. De Sousa 
for useful discussions.
This work was supported by DARPA  and the US ONR.


\begin{references}
\bibitem{long} T$_1$ can be particularly long for bound
states in phosphorus doped silicon~\protect{\cite{feher59,feher61}}.
\end{references}
\end{document}